\documentclass[11pt,twoside]{article}
\usepackage{CAGN2019}
\usepackage{graphicx}

\usepackage[T1]{fontenc} 

\usepackage{latexsym}
\usepackage{verbatim}

\usepackage{ifpdf}  
\ifpdf  
      \DeclareGraphicsExtensions{.pdf,.png,.jpg}  
\else  
      \DeclareGraphicsExtensions{.eps}  
\fi 

\begin{document}

\vskip 1.0cm
\markboth{S. Zamora and A. I. Díaz}{Circumnuclear ionizing clusters}
\pagestyle{myheadings}
%
%
\vspace*{0.5cm}
\parindent 0pt{Contributed  Paper}


\vspace*{0.5cm}
\title{Physical properties of circumnuclear
ionizing clusters: NGC 7742}

\author{S. Zamora $^{1, 2}$, Ángeles I. Díaz,$^{1, 2}$}
\affil{$^{1}$Departamento de Física Teórica, Universidad Autónoma de Madrid (UAM), Spain and $^2$Centro de Investigación Avanzada en Física Fundamental (CIAFF), Spain}

\begin{abstract}
We have analyzed the circumnuclear ring of the spiral galaxy NGC7742 in order to understand its formation and evolution. We have obtained gaseous abundances, characterized the interstellar medium of the clusters and studied the properties of the ionizing clusters. We have also implemented a new methodology using the red wavelength range of optical spectra, with the purpose of understanding how star formation evolves in high metallicity environments.

\bigskip
 \textbf{Key words: } galaxies: abundances --- galaxies: ISM --- techniques: imaging spectroscopy

\end{abstract}

\section{Introduction}
A circumnuclear ring is a region of large gas surface density showing a high rate of star formation. It is usually located within 2 kpc of galactic nuclei and has a stellar mass between 10$^4$M$_\odot$ to 10$^6$M$_\odot$, with hot and massive stars \citep{2011ApJ...727..100W}.These structures can achieve luminosities similar to those of HII regions, contributing significantly to the emission of the whole nuclear region of the galaxy. They are usually formed because gas is accumulated in orbital resonances which are caused by a non-axisymmetric gravitational potential and hence exhibit bars in their morphology \citep{1985A&A...150..327C, 1992MNRAS.259..345A, 1994ApJ...424...84H}.

In this work we have studied NGC7742, which is associated to a group of disk galaxies without stellar bars and without evidence for current interactions with other galaxies. An alternative mechanism for the formation of nuclear rings is a minor merger and this idea has already already been suggested by \citet{2006ApJ...649L..79M} and \citet{2006AJ....131.1336S}. Moreover, its ring is counter-rotating \citep{2002MNRAS.329..513D,2018A&A...612A..66M}, i.e. the gas rotates in the opposite sense to the stellar component. That is why this galaxy constitutes an ideal candidate for such hypothesis and an interesting object of study. In order to understand the effect of the past interaction in its formation and evolution, we have analyzed the properties of its ionized gas and its stellar population. We have focused on the young massive ionizing star clusters with ages under 10 Myr, able to produce HII regions, and stellar masses between 10$^4$ and 10$^6$ M$_\odot$.

\section{Observations and analysis}
\label{procedure}
NGC7742 was observed in the first MUSE Science Verification run (Programme 60.A-9301). MUSE \citep{MUSE} is a integral-field spectrograph (IFS) that covers visible wavelengths, from 4800\r{A} to 9300\r{A}. This spectral range contains hydrogen recombination lines necessary to correct the effects of reddening and forbidden emission lines of [SIII] that allow the gas characterization in terms of electron temperature, density and metal content. On the other hand, the large wavelength range, allows a good determination of colors that are good indicators of the  cluster ages.

\begin{figure}
\begin{center}
\hspace{0.25cm}
\includegraphics[width=0.45\columnwidth]{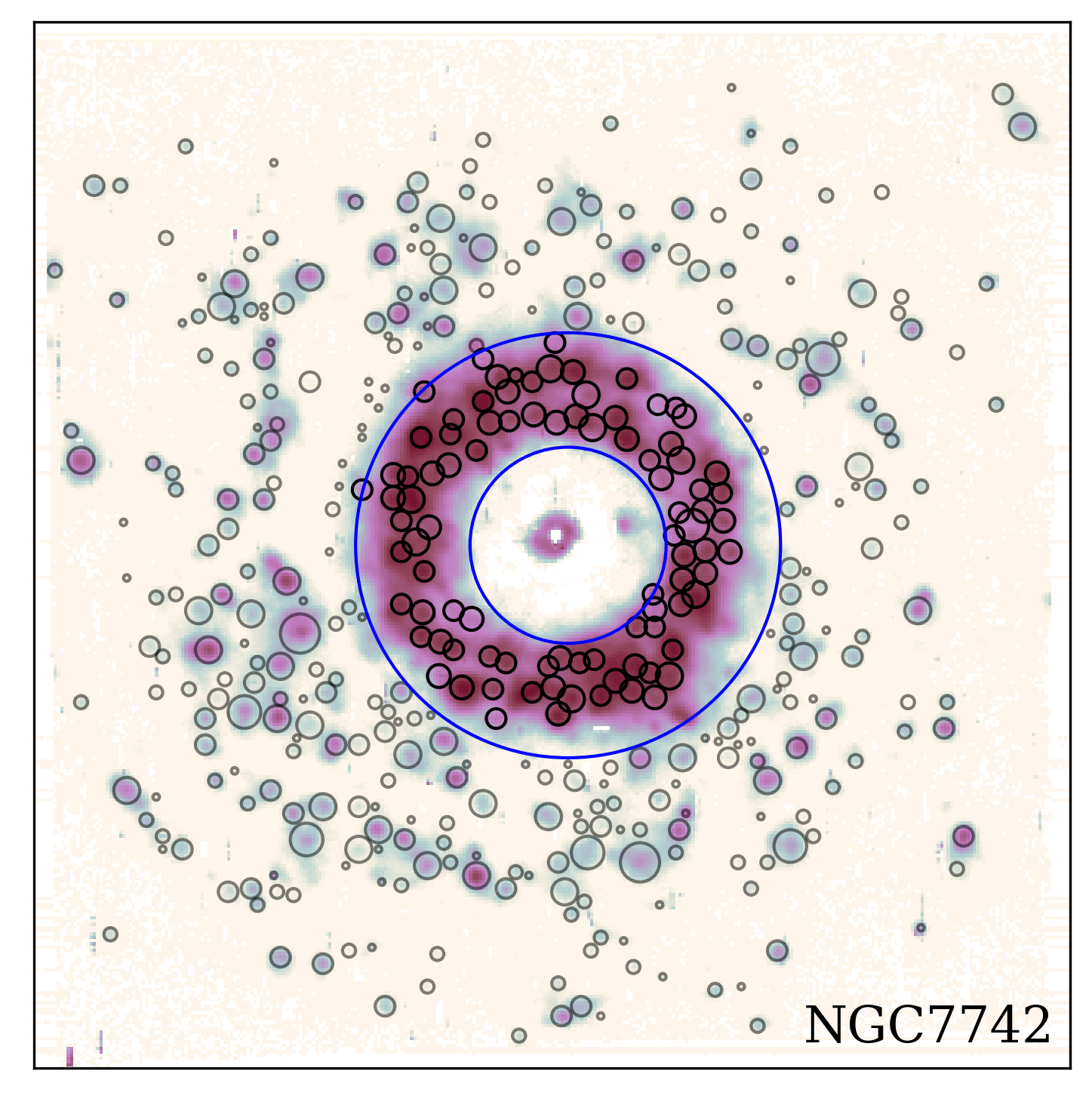}
\caption{HII regions selected using HII$_{EXPLORER}$ on the H$\alpha$ map. Logarithmic color scale.}
\label{Zamora-fig1}
\end{center}
\end{figure}

We have performed an H$\alpha$ map, assuming a linear behavior of the continuum and we have used it to select the star-forming HII regions across the galaxy. Our methodology is based in the philosophy of the HII$_{EXPLORER}$ package \citep{Sanchez2012}, detecting clumps of higher intensity and adding adjacent pixels that follow specific criteria. The ring aperture is based on the H$\alpha$ profile. In addition, we require the following conditions to the integrated spectrum from each selected region: (i) EW(H$\alpha$)>6 \AA\ \citep{CidFernandez2010,Sanchez2015}; (ii) 6>H$\alpha$/H$\beta$>2.7 \citep[][]{Osterbrock2006}. Finally, we have selected a total of 88 regions inside the ring. Additionally, we have analyzed other 158 regions outside the ring as a control sample. Using HST image data (F336W-WFC3) we have confirmed that we are selecting individual clusters with adequate spatial resolution.

We have extracted spectra of each region measuring the intensities of the following strong lines (S/N>3): H$\beta$, H$\alpha$, [OIII]$\lambda\lambda$4959,5007, [NII]$\lambda\lambda $6548,84, [SII]$\lambda\lambda$6716,31 and [SII]$\lambda$9069 ; and the weak lines (S/N>1): [SIII]$\lambda$6312, [OII]$\lambda \lambda $7320,30 (all wavelengths given in \AA ). The fluxes have been corrected for reddening but not for underlying stellar absorption which has been found to amount to less than 3\% of the observed flux. The reddening coefficient has been calculated by adopting the Galactic extinction law of \citet{reddening} (Rv = 2.97, simple screen distribution to the dust).

\section{Results}
\begin{figure}
\begin{center}
\hspace{0.25cm}
\includegraphics[width=\columnwidth]{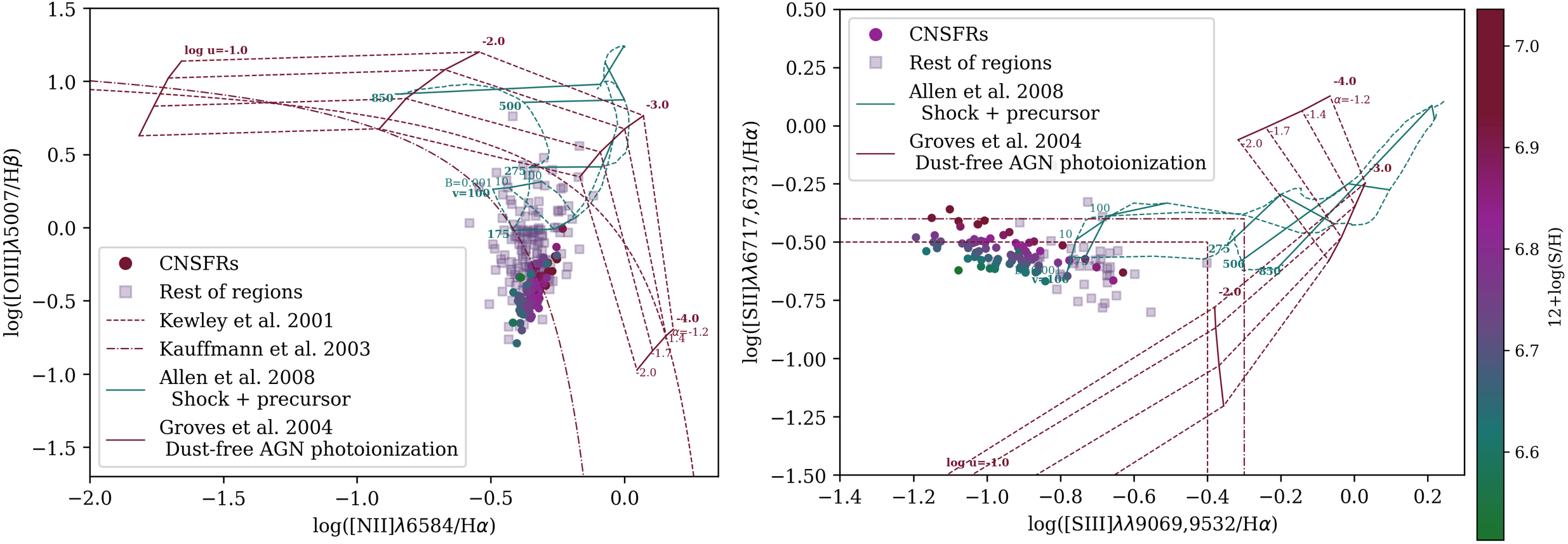}
\caption{Left: The [OIII]/H$\beta$-[NII]/H$\alpha$ diagnostics for emission line objects. Right: The [SII]/H$\alpha$-[SIII]/H$\alpha$ diagnostic diagram.}
\label{Zamora-fig2}
\end{center}
\end{figure}

Left panel of Fig. \ref{Zamora-fig2} shows the distribution of the selected regions in the commonly used, [OIII]/H$\beta$- [NII]/H$\alpha$-diagnostic diagram. According to the classification of \citet{2001ApJ...556..121K}, they are located in the star-forming zone. In the same way, the right panel of the figure shows the sulfur diagnostic diagram, [SIII]/H$\alpha$-[SII]/H$\alpha$. This diagram is more appropriate for our objects since sulfur lines are stronger than those of oxygen at moderate to high metal abundances and it is almost independent of reddening. Using these diagnostics we have discard regions ionized by shocks or AGN.

We have calculated electron temperatures from the ratio of nebular to auroral [SIII] emission line intensities at $\lambda\lambda$ 9069,9532 \AA\ and $\lambda$6312 \AA\ respectively, which have been measured in 6 HII regions \citep[][; T$_e$=5000-25000K, n$_e$=100cm$^{-3}$]{pyneb}:
\begin{equation}
t_{e}([SIII]) = 0.5597-1.808\cdot 10^{-4} R_{S3}+\frac{22.66}{R_{S3}} 
\label{eq:zamora-1}
\end{equation}
\noindent where $t_{e}([SIII]) = 10^{-4} \times T_e([SIII])$. Assuming an only ionization zone in which $T_e(S^+) \approx T_e(S^{++}) = T_e([SIII])$, we have calculated direct ionic abundances using \citep{pyneb}:
\begin{equation}
12+log\left(\frac{S^{+}}{H^{+}}\right)=log\left(\frac{I(6717+6731)}{I(H_{\beta})}\right)+5.516+\frac{0.884}{t_{e}}-0.480\cdot log(t_{e}) 
\label{eq:zamora-2}
\end{equation}
\begin{equation}
12+log\left(\frac{S^{2+}}{H^{+}}\right)=log\left(\frac{I(9069+9532)}{I(H_{\beta})}\right)+6.059+\frac{0.608}{t_{e}}-0.706\cdot log(t_{e}) 
\label{eq:zamora-3}
\end{equation}

For the rest of regions, we have used the S$_{23}$ parameter calibrated against the sulfur abundance, which shows a single valued relation for metallicities up to the solar values:
\begin{equation}
12+log \left(\frac{S^+ + S^{2+}}{H^+}\right)=(6.59\pm 0.01)+ (2.44\pm 0.05)\cdot log S_{23} +(0.91\pm 0.08) \cdot (log S_{23})^2
\label{eq:zamora-4}    
\end{equation}
The resulting sulfur abundances, 12+log(S/H), are between 6.51 and 7.06, about half the solar photospheric value for all the studied regions, hence no corrections for unseen ionization states (namely, S$^{3+}$) are needed.

\begin{figure}
\begin{center}
\hspace{0.25cm}
\includegraphics[width=0.31\columnwidth]{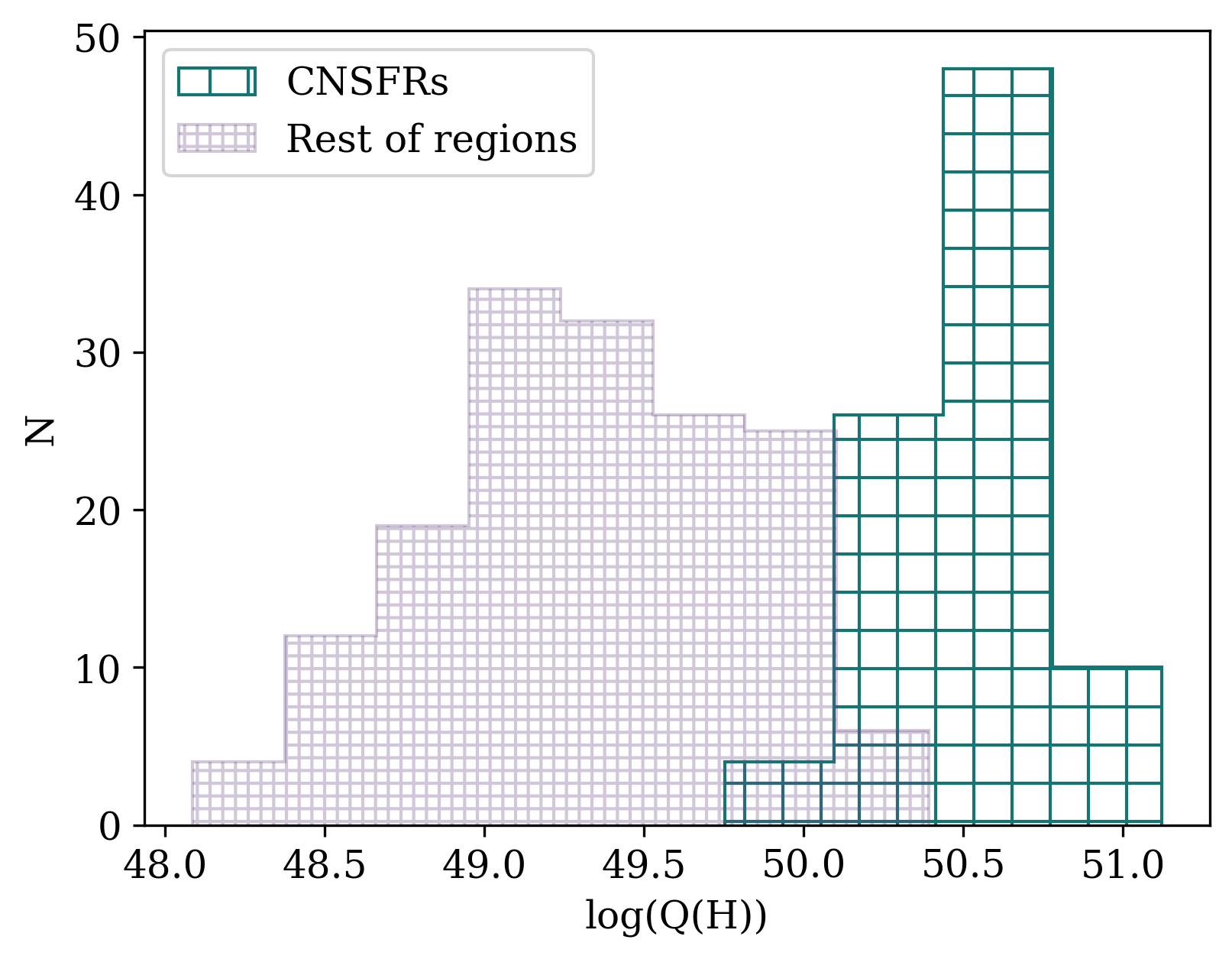}
\includegraphics[width=0.31\columnwidth]{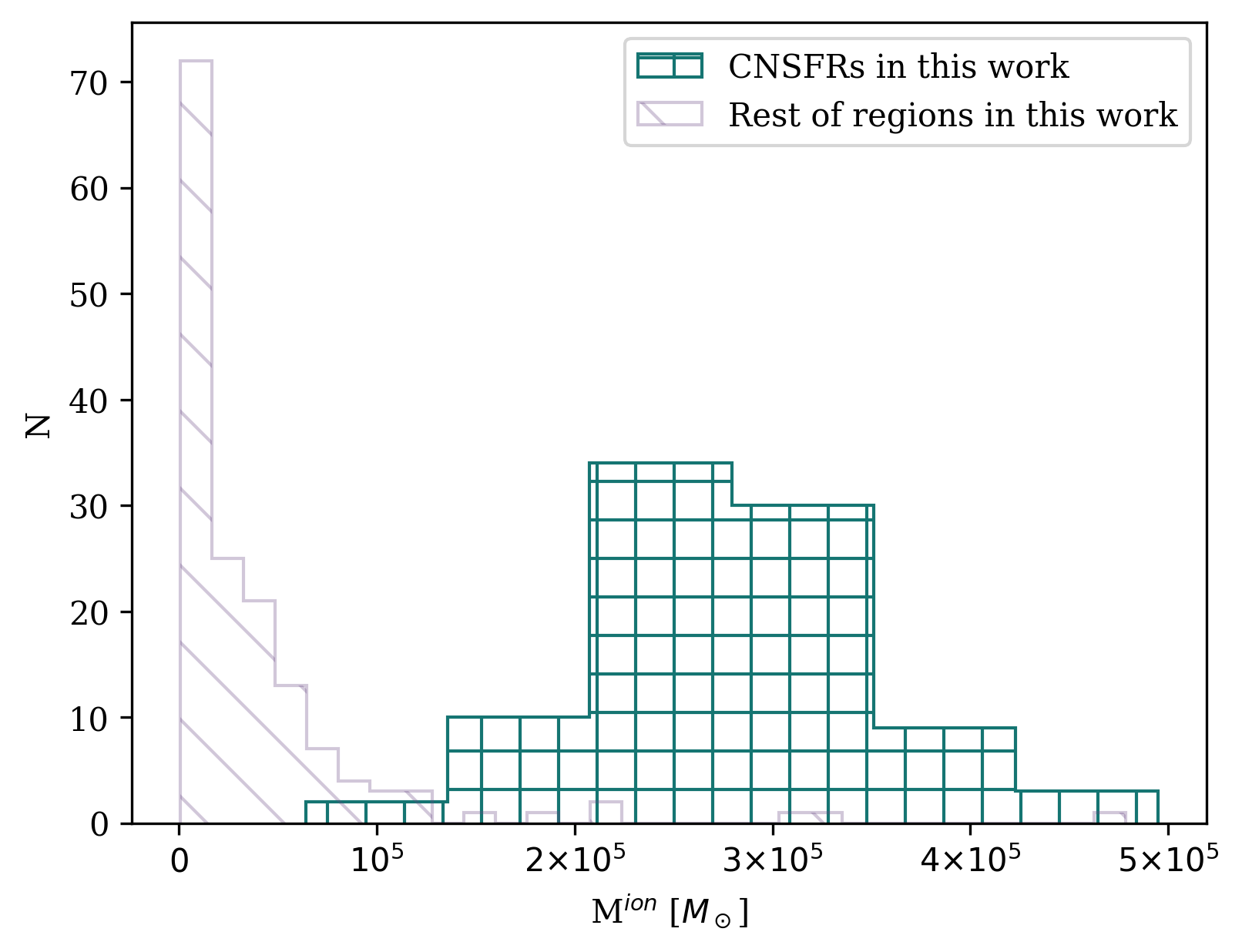}
\includegraphics[width=0.31\columnwidth]{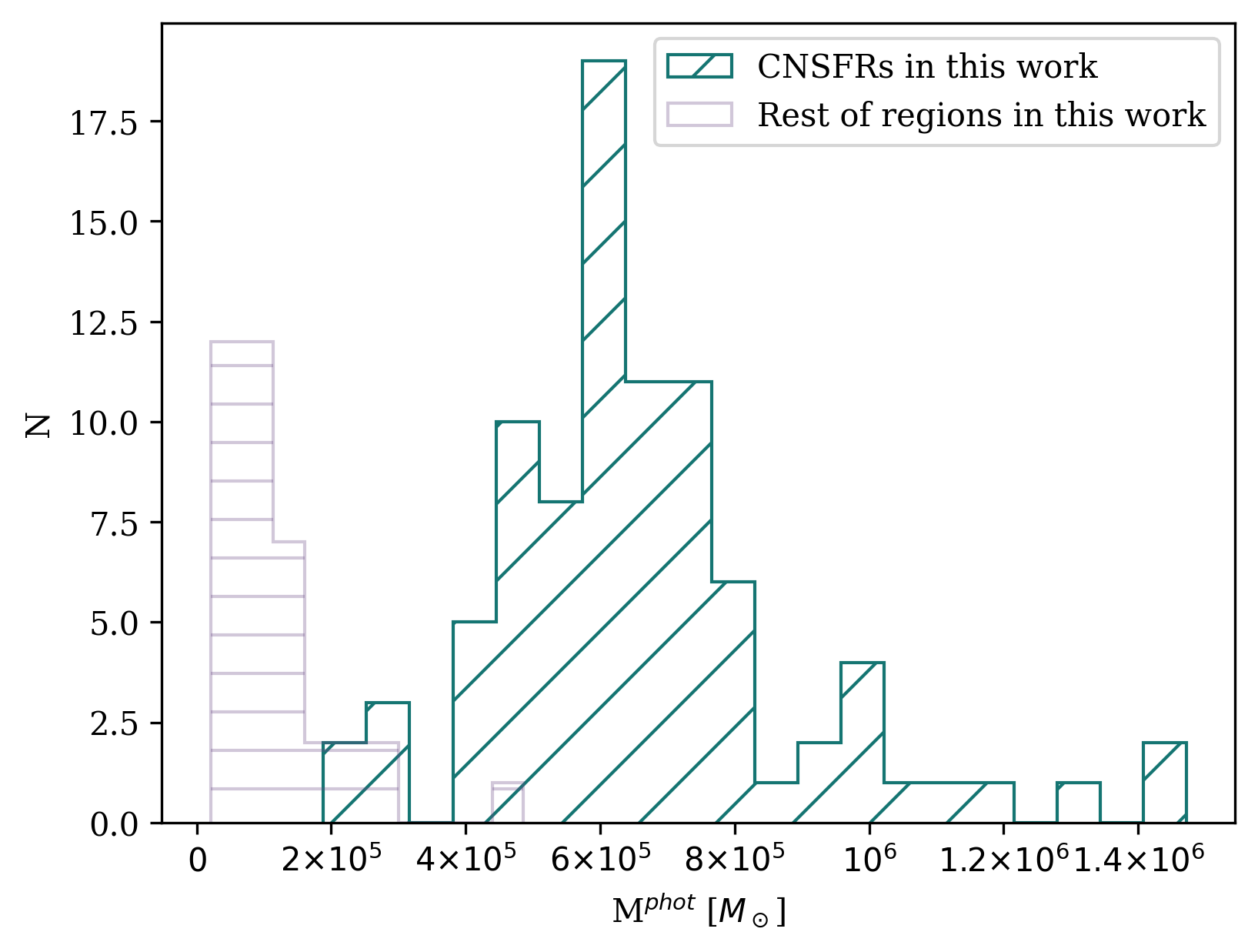}
\caption{Left: Number of ionizing photons. Center: Ionizing masses. Right: Photometric masses.}
\label{Zamora-fig3}
\end{center}
\end{figure}
The number of ionizing photons can be calculated from the reddening corrected H$\alpha$ flux \citep{diaz2000}. The H$\alpha$ luminosities of the individual clusters are 38.11<log(L(H$\alpha$))<39.35, values in the range of those quoted by, for example, \citet{Mazzuca2008}. Ring clusters show higher numbers of ionizing photons, that correspond to larger masses of ionizing clusters under the assumption of no photon escape (see left panel in Fig. \ref{Zamora-fig3}).
The equivalent width of H$\beta$, defined as the intensity of the emission line compared to the continuous emission \citep{Dottori1981}, is an estimator of the age of young stellar populations. For our studied clusters, we find values of log(EW(H$\beta$)) between 0.426 and 1.613. From the number of ionizing photons, the H$\beta$ equivalent widths and metallicity, we can derive their ionizing stellar masses using Simple Stellar Population PopStar models \citep[][; t(Myr)<10, Z=0.004-0.02, Salpeter 1995, m$_{low}$(M$_\odot$) = 0.15 m$_{up}$(M$_\odot$) = 100]{Popstar}. We have obtained and used the following equation:
\begin{equation}
log\left[Q(H)/M_\odot\right] = (44.296\pm0.024)+(0.833\pm0.012)\cdot log\left[EW(H\beta)\right]
\end{equation}

The central panel of Fig. \ref{Zamora-fig3} shows our results (4.806<log(M$^{ion}$)<5.694). These are only lower limits  since we are assuming there is neither dust absorption and re-emission at IR wavelengths or photon escape from HII regions.

Magnitudes and colors have been calculated from their spectra as a first approach to characterize the properties of the stellar populations. We have used the expressions from \citet{Fukugita1995}. The color is very similar in both samples, within the ring and out of it, but, systematically, regions outside the ring have higher magnitudes (less luminosity). From the I-band and R-band magnitudes, H$\beta$ equivalent widths, and metallicity, we can derive the photometric masses of the ionizing clusters using \citet{Popstar} models. The estimation of these masses provides information about the average evolutionary stage of them and they are found to be between 5.3<log M$^{phot}$<6.2 (see right panel of Fig.\ref{Zamora-fig3})

There is an intrinsic difference between photometric and ionizing masses of the clusters in this study, with the first being  more than twice the former. This ratio seems to be constant for the whole sample (M$^{phot}$/M$^{ion}$ = 2.4).

\begin{figure}
\begin{center}
\hspace{0.25cm}
\includegraphics[width=0.48\columnwidth]{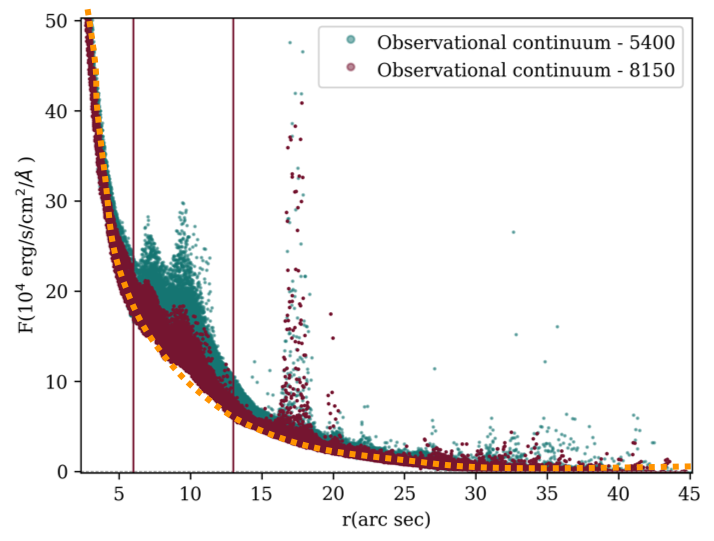}
\includegraphics[width=0.48\columnwidth]{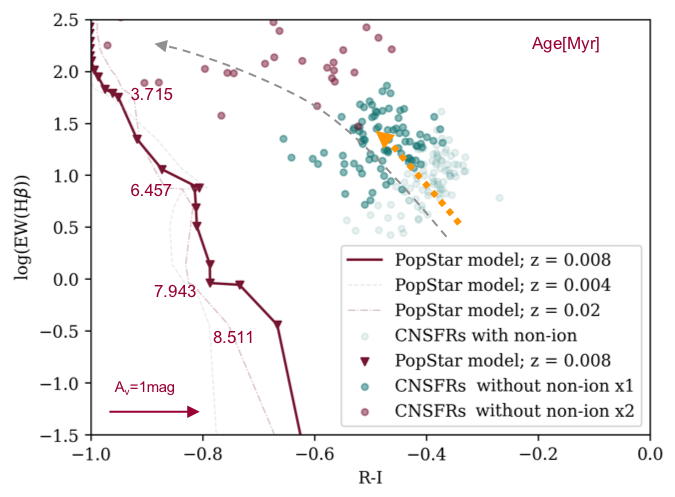}
\caption{Left: Continuum flux of individual spaxels at $\lambda$ 5400 \AA\ (blue) and $\lambda$ 8150 \AA\ (red) as a function of galaxy radius. Limits of the ring shown in vertical red lines. Right: The relation between the equivalent width of H$\beta$ and the (R-I) color for individual clusters (open green circles). The yellow arrow shows the change of position of these clusters once the underlying disk stellar population has been subtracted once (filled green circles) and twice (filled red circles).}
\label{Zamora-fig4}
\end{center}
\end{figure}

To check for the presence of a non-ionizing population, we have produced a combined spectrum of an annular region surrounding the ring, at R =1.63 Kpc, i.e., external to it. It represents a lower limit to the underlying population from the galaxy disk. A similar annulus taken at the inner side of the ring shows twice the amount of continuum light. We have subtracted this representative spectrum from each region, and we have calculated again the H$\beta$ equivalent widths and (R-I) colors. 

The left panel of Fig. \ref{Zamora-fig4} shows pixel-to-pixel continuum fluxes at the blue and red spectral windows. In the ring, a blue excess, with respect to the galaxy profile, is evident and can be identified with a young stellar population. A red population can also be identified underlying this excess, that can be associated with the presence of red supergiant stars. In fact, the prominent Ca triplet lines seen in the spectra point to that. The panel at the right of the figure shows the H$\beta$ equivalent with vs the (R-I) color. The first is an indicator of the ionizing stellar population age (<10 Ma) and the second is an indicator of the age of any non-ionizing population. The position of the individual clusters in this diagram, once the underlying disk stellar population has been subtracted, indicates a narrow range of ages, between 5.5 and 7.5 Ma. Moreover, we can still see an excess of color, larger values of R-I, which suggests the presence of ionizing and non-ionizing composite stellar populations. 

\section{Conclusions}
\label{discussion}
We can conclude that young stars in the ring of NGC7742 have been formed, almost simultaneously, and they have an age younger than 7.5 Ma. In most cases an underlying stellar population can be identified, besides that of the galaxy disk, pointing to a composite population in the ring. 

\acknowledgments This research has made use of the services of the ESO Science Archive Facility. 
and it has been supported by Spanish grants from the  former Ministry of Economy, Industry and Competitiveness through the MINECO-FEDER research grant AYA2016-79724-C4-1-P, and the associated contract BES-2017-080509. 

\bibliographystyle{aaabib}
\bibliography{Zamora}

\end{document}